\documentclass[aps,pre,reprint,superscriptaddress,longbibliography,floatfix]{revtex4-2}
\usepackage{amsmath,amssymb,bm,graphicx,microtype}
\usepackage[colorlinks=true,linkcolor=blue,citecolor=blue,urlcolor=blue]{hyperref}
\newcommand{\bs}{\bm\sigma}
\newcommand{\Inv}{\mathcal I}
\newcommand{\Fix}{\operatorname{Fix}}
\newcommand{\dIsing}{\texorpdfstring{$\delta$}{delta}-Ising}

\begin{document}

\title{Kinetic Sectors and Nonequilibrium Dynamics in the \texorpdfstring{$\delta$}{delta}-Ising Model}

\author{Dehen Cao}
\affiliation{School of Physics, Huazhong University of Science and Technology, Wuhan 430074, China}%

\author{Wei Jiang}
\email{weijiang@hust.edu.cn}
\affiliation{School of Physics, Huazhong University of Science and Technology, Wuhan 430074, China}%
\affiliation{International Joint Research Laboratory of Magnetic Confinement Fusion and Plasma Physics, State Key Laboratory of Advanced Electromagnetic Engineering and Technology, School of Electrical and Electronic Engineering, Huazhong University of Science and Technology, Wuhan 430074, China}%

\author{Ya Zhang}
\email{yazhang@whut.edu.cn}
\affiliation{School of Physics and Mechanics, Wuhan University of Technology, Wuhan 430070, China}%

\date{\today}
\begin{abstract}
We introduce and analyze the $\delta$-Ising model, a kinetic Ising ferromagnet in which single-spin flips are accepted when their energy change does not exceed a tunable threshold. Varying the field and threshold opens or closes local flip channels, dividing parameter space into kinetic sectors, each governed by a fixed Markov transition operator. We prove that the dynamics retains global spin-inversion symmetry exactly when up and down spins have equal numbers of open channels. In one such sector on a periodic chain of length $L$, unbiased random initial states therefore reach the field-aligned and antifield states with probability $1/2$ each. Closing a single channel leaves both states absorbing but makes the probability of ordering against the field decay as $(2/\pi)^L$. Parameter sweeps correspond to ordered products of sector operators, which separate reaching a state from holding it, so hysteresis appears when a state is reached and released at different sector boundaries. We also show how increasing the system size can induce a Mpemba-like reversal, allowing a state with more misaligned spins to reach the final ordered state sooner on average. On a periodic chain, this reversal arises because the mean absorption time grows logarithmically with system size for many short domains but linearly for a single macroscopic domain, and it disappears when the bulk flip channel is opened. These findings show how a simple, intuitive rule for Ising dynamics links local kinetic constraints to field response, memory, and relaxation within a unified framework.
\end{abstract}
\maketitle

\section{Introduction}\label{sec:intro}

The Ising model provides a simple framework for studying equilibrium phase transitions and nonequilibrium relaxation~\cite{Ising1925,Onsager1944,Glauber1963,Bray1994}. On a regular graph of degree $z$, single-spin-flip Ising dynamics with uniform couplings and a uniform field has $2(z+1)$ local flip channels, each defined by the spin orientation and the number of neighbors aligned with it~\cite{Krapivsky2010}. At finite temperature, all flip channels remain open under Glauber and Metropolis dynamics, so varying the temperature or field changes only their rates~\cite{Glauber1963,Metropolis1953}. At zero temperature, a hard cutoff instead closes all channels that raise the energy~\cite{NewmanStein2000,Spirin2001a}. Evolving through the remaining channels, the system can freeze into stripe states on a square lattice or become trapped in blinker states on a cubic lattice~\cite{Spirin2001b,Olejarz2011}. Yet the zero-temperature cutoff allows only a small subset of the possible combinations of the $2(z+1)$ flip channels. Whether other combinations yield distinct physical behavior remains an open question.

Models with flip rules based on neighbor counts support this possibility, showing how local constraints alone can generate complex collective behavior. For example, freezing spins of one orientation and allowing the others to flip only when they have at most a threshold number of aligned neighbors yields bootstrap percolation, which can undergo a discontinuous transition on the Bethe lattice~\cite{Chalupa1979}. In the Fredrickson--Andersen model, requiring enough excited neighbors for a spin to flip produces glassy slowing down~\cite{Fredrickson1984,Fredrickson1985} and spatially heterogeneous relaxation~\cite{GarrahanChandler2002}, even without energetic interactions between spins. For its two-spin facilitated version, recent work links the first excitation time at a site to the threshold constant of two-neighbor bootstrap percolation in the low-density limit~\cite{Hartarsky2023}. In generalized $\mathcal U$-Ising dynamics, neighbor subsets define the allowed flips, and the system can exhibit complete fixation or persistent flipping at some sites when frozen spins are present~\cite{Mareche2025}. Similarly, in zero-temperature Ising dynamics, the energy change upon flipping increases with the number of aligned neighbors for a given spin orientation, so rejecting energy-increasing flips imposes a threshold on this number. In a weak field, zero-temperature Ising dynamics has been related to bootstrap percolation~\cite{Spirin2001b}. The zero-temperature response of the random-field Ising model to a monotonically increasing field has also been linked to bootstrap percolation with blocked sites~\cite{Ding2026}. At zero temperature, the random-field Ising model exhibits hysteresis and return-point memory under quasistatic field cycling~\cite{Sethna1993}. Ising ferromagnets also show a magnetic counterpart of the Mpemba effect, ordering faster from hotter, less correlated initial states after a quench to zero temperature~\cite{Vadakkayil2021}. These examples illustrate how different local flip rules give rise to diverse collective behavior, and the zero-temperature results further show that hysteresis and anomalous relaxation can arise without thermal activation. Together, these findings suggest that tuning such rules within a single Ising model may reveal a similarly rich range of phenomena.

To test this possibility, we introduce the $\delta$-Ising model, a kinetic Ising ferromagnet in which a proposed single-spin flip is accepted if and only if its energy change satisfies $\Delta H\leq E_0$, as in threshold-accepting optimization~\cite{Dueck1990}. Setting $E_0=0$ recovers zero-temperature Metropolis dynamics, while varying $E_0$ changes which spin flips are allowed without altering the Ising Hamiltonian or its field coupling. On a $z$-regular graph, a spin's orientation and number of aligned neighbors define a flip channel that is either open or closed. The channel states change only across two families of parallel threshold lines, which divide the $(h,E_0)$ plane into $(z+2)^2$ sectors with fixed dynamics. Within each sector, a fixed Markov transition operator governs the dynamics, and local neighbor counts determine its fixed configurations. The diagonal sectors recover symmetric threshold-voter rules~\cite{Liggett1994,DurrettSteif1993}, whereas off-diagonal sectors yield asymmetric, field-biased dynamics. This finite table of sector operators describes relaxation at fixed parameters, while ordered products of these operators describe multistep protocols.

Combining exact analysis of regular graphs and periodic chains with Monte Carlo simulations in one to three dimensions, we obtain the following three main results. (i)~When up and down spins have equal numbers of open channels, the sector's transition operator equals its zero-field counterpart and commutes with global spin inversion. This field screening preserves zero-field statistics at nonzero fields, including equal probabilities of $1/2$ for the two ordered states on a periodic chain with unbiased initial conditions, a stripe probability of about $1/3$ in two dimensions~\cite{Barros2009,Olejarz2012}, and persistent blinker activity with fluctuating total energy in three dimensions~\cite{Spirin2001b,Olejarz2011}. Closing a single channel on the periodic chain preserves both ordered fixed points, but the probability of reaching the antifield state then decays as $(2/\pi)^L$ with chain length $L$. (ii)~During round-trip parameter sweeps, a written state can remain fixed in later sectors where it would rarely be reached. When writing and release occur at different sector boundaries, this persistence produces hysteresis. In two dimensions, identical zero-field threshold sweeps with finite dwell times produce hysteresis with probability about $2/3$, while the remaining $1/3$ of trajectories freeze into the stripe states of (i) without hysteresis. (iii)~When the target operator allows boundary flips but blocks bulk flips, interior spins must wait for flips to propagate inward, in an order constrained by a bootstrap-percolation closure. Many thin reversed domains can therefore disappear faster on average than a single thick domain containing fewer reversed spins, producing a Mpemba-like relaxation reversal. In one dimension, their mean absorption times scale as $\ln L$ and $L$, respectively, so increasing system size induces this reversal. However, crossing a sector boundary to open the single bulk flip channel eliminates the reversal in one dimension and largely suppresses it in two dimensions.

\section{Model and Operator Framework}\label{sec:model}

\subsection{The \dIsing\ model}\label{sec:model-def}

Consider a finite connected $z$-regular graph $G=(V,E)$ with $N=|V|$ vertices. An Ising spin $\sigma_i=\pm1$ sits on each vertex $i\in V$, the configuration space is $\Omega_G=\{-1,+1\}^{V}$, and the Hamiltonian is
\begin{equation}
H(\bs)=-J\sum_{\langle ij\rangle}\sigma_i\sigma_j-h\sum_i\sigma_i,
\label{eq:H}
\end{equation}
where $J>0$ is the ferromagnetic coupling and $h$ is the external field. The dynamics uses random asynchronous single-spin flips. Each attempt selects a vertex $i$ uniformly and proposes $\sigma_i\to-\sigma_i$. If vertex $i$ has $k\in\{0,1,\ldots,z\}$ aligned neighbors before the flip, the energy change of the flip is
\begin{equation}
\Delta H_i = 2J(2k-z)+2h\sigma_i.
\label{eq:dH}
\end{equation}
Each pair $(k,\sigma_i)$ defines a local flip channel. The $z+1$ possible values of $k$ and the two orientations of $\sigma_i$ give $2(z+1)$ channels. On a periodic $d$-dimensional hypercubic lattice of linear size $L$, $z=2d$ and $N=L^d$, so Eq.~\eqref{eq:dH} reduces to $\Delta H_i=4J(k-d)+2h\sigma_i$.

Every local flip channel is judged by the same threshold $E_0$. A flip is accepted if and only if $\Delta H_i\leq E_0$, that is,
\begin{equation}
A_{E_0}(\Delta H_i) = \Theta(E_0 - \Delta H_i),
\label{eq:accept}
\end{equation}
with $\Theta(0)=1$, so that each channel is either open or closed. We call this model the $\delta$-Ising model. The name reflects the fixed threshold $\varepsilon=E_0$, described by $p(\varepsilon)=\delta(\varepsilon-E_0)$, where $\delta$ is the Dirac delta distribution. For $E_0=-2H_c<0$, the acceptance condition becomes $\sigma_i h_i^{\mathrm{loc}}\leq-H_c$, where $h_i^{\mathrm{loc}}=J\sum_{j\sim i}\sigma_j+h$ is the local field. This is the athermal switching criterion for interacting hysterons with a uniform coercive field $H_c$~\cite{Teunisse2025}. With $E_0$ fixed, the transition probabilities depend only on the current configuration, so the dynamics defines a Markov chain on $\Omega_G$. Let $\bs^{(i)}$ denote the configuration obtained by flipping spin $i$. Uniform site selection gives the single-attempt transition kernel
\begin{equation}
\begin{aligned}
P_{h,E_0}(\bs,\bs^{(i)})
&=\frac{1}{N}\Theta(E_0-\Delta H_i),\\
P_{h,E_0}(\bs,\bs)
&=1-\frac{1}{N}\sum_{i=1}^{N}\Theta(E_0-\Delta H_i).
\end{aligned}
\label{eq:kernel}
\end{equation}
All other matrix elements vanish. Each nonzero off-diagonal element of $P_{h,E_0}$ represents a single-spin flip allowed by the threshold. These transitions define a directed graph on $\Omega_G$. A probability row vector $\mu$ on $\Omega_G$ evolves to $\mu P_{h,E_0}$ after one attempt. One Monte Carlo sweep consists of $N$ attempts, with transition operator
\begin{equation}
Q_{h,E_0}=P_{h,E_0}^{N},
\label{eq:Q}
\end{equation}
so the distribution after one sweep is $\mu Q_{h,E_0}$. Below, we use $P$ for single attempts and the transition graph, and $Q$ for propagating distributions and composing operators over sweeps. For simplicity, we express $H$, $h$, and $E_0$ in units of $J$. Temperature $T$ is measured in units of $J/k_{\mathrm B}$.

\subsection{Kinetic sectors and the finite operator table}\label{sec:sectors}

\begin{figure}[t]
\centering
\includegraphics[width=\columnwidth]{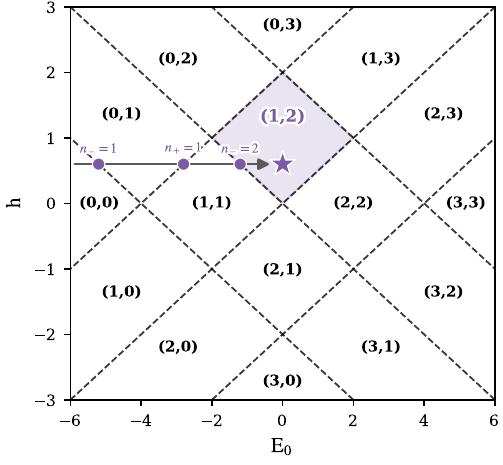}
\caption{Kinetic sectors of a one-dimensional chain and their $(n_+,n_-)$ labels. The six threshold lines of Eq.~\eqref{eq:thresholds} divide the $(E_0,h)$ plane into 16 sectors, each labeled by its numbers of open channels $(n_+,n_-)$ for up and down spins. The horizontal arrow is a path of increasing $E_0$ at fixed $h$, and the three dots on it mark where $n_-$ rises to 1, $n_+$ to 1, and $n_-$ to 2. The star at $E_0=0$ lies in the sampled sector $(1,2)$.}
\label{fig:sectors}
\end{figure}

The step function in Eq.~\eqref{eq:accept} makes each local channel $(k,\sigma)$ either open or closed. A channel switches at $E_0=\Delta H(k,\sigma)$, giving the threshold lines
\begin{equation}
E_0=2(2k-z)+2h\sigma.
\label{eq:thresholds}
\end{equation}
In the $(E_0,h)$ plane, these form two families of $z+1$ parallel lines, one for each spin orientation $\sigma=\pm1$. In the $h(E_0)$ representation used in Fig.~\ref{fig:sectors}, the two families have slopes $+1/2$ and $-1/2$. Each line in one family therefore intersects every line in the other, partitioning the parameter plane into $(z+2)^2$ kinetic sectors. The acceptance pattern is constant within each sector, so the single-attempt kernel $P$ and sweep operator $Q$ remain unchanged. Crossing a threshold line opens or closes the corresponding class of channels and adds or removes the associated transition edges.

Sectors are thus distinguished by which local channels are open. For each spin orientation $\sigma$, we define the number of open channels as
\begin{equation}
n_\sigma=\#\bigl\{k:\Delta H(k,\sigma)\leq E_0\bigr\}
\in\{0,1,\ldots,z+1\},
\label{eq:nsigma}
\end{equation}
where $\#$ counts the allowed values of $k$. Since $\Delta H(k,\sigma)$ increases strictly with $k$, the open channels are precisely those with $k<n_\sigma$. The pair $(n_+,n_-)$ therefore uniquely specifies the open channels and the sector's transition rules. Writing $u_i=z-k_i$ and $\theta_\sigma=z-n_\sigma+1$, the acceptance condition becomes $u_i\geq\theta_{\sigma_i}$. With independent unit-rate Poisson clocks at the sites, a diagonal sector $n_+=n_-=n$, with $1\leq n\leq z$, has flip rates $c_i(\bs)=\mathbf{1}\{u_i\geq z-n+1\}$, giving the symmetric threshold voter model~\cite{DurrettSteif1993}. Figure~\ref{fig:sectors} illustrates this labeling in one dimension. Starting from sector $(0,0)$, where all channels are closed, we increase $E_0$ at fixed $h$. Each threshold crossing for orientation $\sigma$ opens one channel type and increases $n_\sigma$ by one. The path reaches the starred sector after crossing one threshold line for $+1$ spins and two for $-1$ spins, so this sector is labeled $(1,2)$. Here, $+1$ spins can flip only for $k=0$, whereas $-1$ spins can flip for $k=0$ or $1$. We can therefore label sectors by counting threshold crossings, without computing trajectories.

The label $(a,b)$ specifies the open channels in a sector. For a fixed graph $G$, it therefore uniquely determines the sweep operator, which is constant throughout the sector. Denoting this sector by $\mathcal S_{a,b}$, we define
\begin{equation}
Q^{(G)}(a,b)\equiv Q_{h,E_0},
\qquad (h,E_0)\in\mathcal S_{a,b}.
\label{eq:table}
\end{equation}
The operators $Q^{(G)}(a,b)$ with $0\leq a,b\leq z+1$ form a finite operator table. For a fixed graph $G$, this table represents the continuous $(h,E_0)$ plane exactly by $(z+2)^2$ discrete entries. The table size depends only on the coordination number $z$, with 16 and 36 entries for one- and two-dimensional hypercubic lattices, respectively. The operators themselves depend on the underlying graph $G$. For a periodic $d$-dimensional hypercubic lattice, we write $Q^{(d)}(a,b)$. For brevity, we denote a generic operator by $Q$.

\subsection{Operator fixed points}\label{sec:fixed-points}

For any operator $Q$ in the finite operator table, a configuration $\bs$ is a configurational fixed point if one sweep leaves it unchanged with probability $1$. We denote the set of all such configurations by $\Fix(Q)$. This is equivalent to the absence of open flip channels in the configuration. By Sec.~\ref{sec:sectors}, the channel of site $i$ is open when $k_i<n_{\sigma_i}$, so
\begin{equation}
\Fix(Q^{(G)}(n_+,n_-))
=
\left\{
\bs\in\Omega_G:
k_i(\bs)\geq n_{\sigma_i}
\ \text{for all }i
\right\}.
\label{eq:fix}
\end{equation}
Equation~\eqref{eq:fix} also determines how the fixed-point sets of different operators are nested. If $n'_+\geq n_+$ and $n'_-\geq n_-$, then
\begin{equation}
\Fix(Q^{(G)}(n'_+,n'_-))
\subseteq
\Fix(Q^{(G)}(n_+,n_-)).
\label{eq:nest}
\end{equation}
Thus, opening more local channels can remove fixed configurations but cannot create new ones. At the two extremes, $\Fix(Q^{(G)}(0,0))=\Omega_G$ and $\Fix(Q^{(G)}(z+1,z+1))=\varnothing$. In a uniform state, $k_i=z$ at every site, so the all-up and all-down states are fixed points if and only if $n_+\leq z$ and $n_-\leq z$, respectively. When $n_-=z+1$, no fixed configuration can contain down spins, so the all-up state is the unique fixed point for $n_+\leq z$. Exchanging the signs gives the corresponding result on the opposite boundary.

\begin{table}[b]
\caption{\label{tab:fix}Fixed-point sets of the 16 sector operators $Q^{(1)}(n_+,n_-)$ on the one-dimensional periodic chain. Rows and columns correspond to $n_+$ and $n_-$. $\bs_+$ and $\bs_-$ denote the all-up and all-down states, and $\mathcal H=\{\bs_+,\bs_-\}$ is the set formed by the two uniform states. $\mathcal D_\sigma$ denotes the set of configurations without isolated $\sigma$ spins, and $\mathcal D=\mathcal D_+\cap\mathcal D_-$ is the set of configurations without any isolated spin. All entries follow from Eq.~\eqref{eq:fix}.}
\begin{ruledtabular}
\begin{tabular}{ccccc}
$n_+\backslash n_-$ & $0$ & $1$ & $2$ & $3$\\
\hline
$0$ & $\Omega_G$ & $\mathcal D_-$ & $\mathcal H$ & $\{\bs_+\}$\\
$1$ & $\mathcal D_+$ & $\mathcal D$ & $\mathcal H$ & $\{\bs_+\}$\\
$2$ & $\mathcal H$ & $\mathcal H$ & $\mathcal H$ & $\{\bs_+\}$\\
$3$ & $\{\bs_-\}$ & $\{\bs_-\}$ & $\{\bs_-\}$ & $\varnothing$\\
\end{tabular}
\end{ruledtabular}
\end{table}

The one-dimensional periodic chain provides the simplest complete example of this nesting. Table~\ref{tab:fix} lists the fixed-point sets of its 16 sector operators. At $(n_+,n_-)=(0,0)$, all flip channels are closed and every configuration is fixed. Raising either coordinate to $1$ excludes configurations containing isolated spins of that sign, giving $\mathcal D_+$ or $\mathcal D_-$. Raising both to $1$ gives their intersection $\mathcal D$, in which every spin domain has length at least two. For $n_+,n_-\leq2$, setting either coordinate to $2$ excludes all mixed configurations because each violates Eq.~\eqref{eq:fix} at a domain boundary. Only the two uniform states in $\mathcal H$ remain. If one coordinate reaches $3$ while the other remains below $3$, no fixed configuration can contain spins of the corresponding sign, leaving only the opposite uniform state. At $(3,3)$, the fixed-point set is empty. Thus, moving rightward or downward in the table can only reduce the fixed-point set, showing how local flip channels constrain the global fixed-point structure.

\subsection{Observables}\label{sec:observables}

To compare the dynamical responses of different sectors, we use four observables, namely the magnetization $M$, the largest-cluster fraction $f_{\max}$, the temporal standard deviation of the magnetization $\sigma_M$, and the energy fraction $\mathcal R$.

The magnetization, which measures the imbalance between the two spin orientations, is defined as
\begin{equation}
M(\bs)=N^{-1}\sum_{i=1}^{N}\sigma_i.
\label{eq:M}
\end{equation}
It lies in $[-1,1]$, with its sign indicating the majority spin orientation and $|M|$ measuring the degree of polarization. Below, $\langle\cdot\rangle$ denotes an ensemble average and $\langle\cdot\rangle_t$ denotes a time average along a single trajectory over a specified window.

The same magnetization can arise from different domain structures. To characterize the connectivity of aligned spins, we define the largest-cluster fraction
\begin{equation}
f_{\max}(\bs)=N^{-1}\max_{c\in\mathcal C(\bs)}|c|.
\label{eq:fmax}
\end{equation}
Here $\mathcal C(\bs)$ is the set of clusters formed by nearest-neighbor connections between aligned spins, including connections across periodic boundaries. This fraction lies in $[N^{-1},1]$ and is invariant under global spin inversion, so it is unaffected by the cancellation of magnetization in symmetric ensembles. Its scaling with system size distinguishes macroscopic clusters occupying a finite volume fraction from fragmented structures.

Trajectories with the same time-averaged magnetization can differ in fluctuation amplitude. We quantify these fluctuations within the sampling window by the temporal standard deviation along each trajectory,
\begin{equation}
\sigma_M=\sqrt{\left\langle M^{2}\right\rangle_t-\left\langle M\right\rangle_t^{2}}.
\label{eq:sigmaM}
\end{equation}
A nonzero $\sigma_M$ indicates magnetization fluctuations within the window. Once the system reaches a fixed point, $\sigma_M=0$ in all subsequent windows.

To compare energies across dimensions and fields, we use the mean energy of unbiased Bernoulli initial states and the field-aligned ground-state energy as references. On a $z$-regular graph, these are $\langle H\rangle_{\mathrm{init}}=0$ and $H_{\mathrm{gs}}=-N(z/2+|h|)$, respectively. We then define the fraction of unreleased available energy as
\begin{equation}
\mathcal{R}=\frac{\langle H\rangle-H_{\mathrm{gs}}}{\langle H\rangle_{\mathrm{init}}-H_{\mathrm{gs}}}
=\frac{\langle H\rangle}{|H_{\mathrm{gs}}|}+1.
\label{eq:R}
\end{equation}
Here $\mathcal R=0$ corresponds to the ground state and $\mathcal R=1$ to the initial ensemble mean, while $\mathcal R>1$ indicates a mean energy above this initial value. The energy fraction $\mathcal R$ can vary with $h$ even within a sector where the configuration dynamics is unchanged.

\section{Dynamics under Individual Sector Operators}\label{sec:single}

\begin{figure*}[t]
\centering
\includegraphics[width=0.92\textwidth]{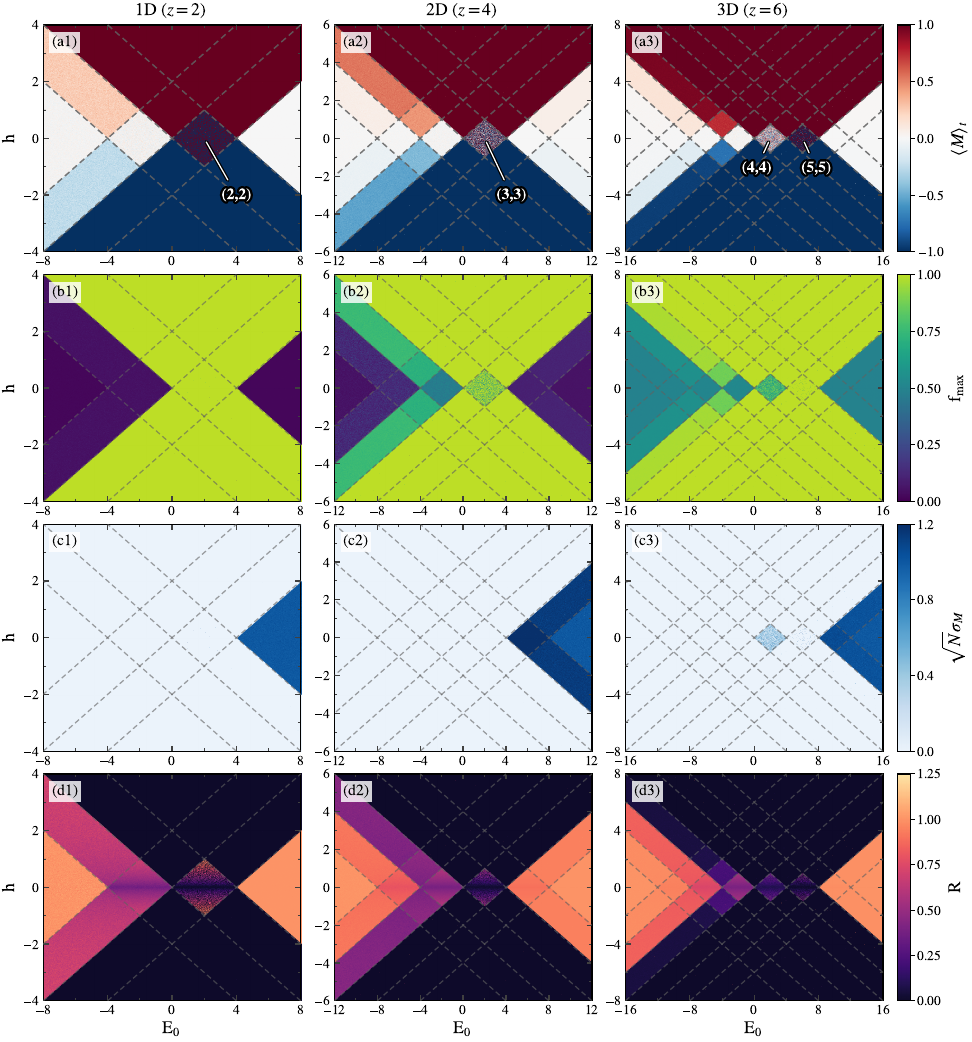}
\caption{Late-time observables in the $(E_0,h)$ plane for periodic lattices in one to three dimensions. (a1)--(a3) Time-averaged magnetization $\langle M\rangle_t$. (b1)--(b3) Largest-cluster fraction $f_{\max}$ of aligned spins. (c1)--(c3) Temporal standard deviation of the magnetization, shown as $\sqrt N\sigma_M$. (d1)--(d3) Fraction of unreleased available energy $\mathcal{R}$. Dashed lines are the analytic threshold lines of Eq.~\eqref{eq:thresholds}, and labels in (a1)--(a3) give sector indices $(n_+,n_-)$. The chain, square lattice, and cubic lattice have $L=400$, $64$, and $24$, and each map is computed on an $800\times800$ grid over the plotted range. Each point comes from a single trajectory started from an independent Bernoulli initial state, with observables sampled over 400 sweeps following $199\,600$ sweeps of evolution in all three dimensions.}
\label{fig:sector-observables}
\end{figure*}

Section~\ref{sec:model} mapped the continuous parameter plane onto a finite operator table according to which local flip channels are open. We now use the parameter scans in Fig.~\ref{fig:sector-observables} to examine the late-time macroscopic behavior generated by these operators. The three columns show results in one, two, and three dimensions, from left to right. From top to bottom, the four rows show the magnetization, the largest-cluster fraction, the temporal standard deviation of the magnetization, and the fraction of unreleased available energy, defined in Eqs.~\eqref{eq:M}--\eqref{eq:R}. The main boundaries between response regions coincide with the analytic threshold lines of Eq.~\eqref{eq:thresholds}. Within individual sectors, however, the observables vary across parameter points, each representing a single independent trajectory. 

Figure~\ref{fig:sector-observables} shows several unusual features. In the one-dimensional $Q^{(1)}(2,2)$ sector, the magnetization takes either sign across trajectories, while the largest-cluster fraction is always unity [Figs.~\ref{fig:sector-observables}(a1) and \ref{fig:sector-observables}(b1)]. Together, these observations indicate that different trajectories reach uniform terminal states with opposite orientations. In the two-dimensional $Q^{(2)}(3,3)$ sector, the largest-cluster fraction instead varies from trajectory to trajectory [Fig.~\ref{fig:sector-observables}(b2)], reflecting differences in domain structure. In the three-dimensional $Q^{(3)}(4,4)$ sector, some trajectories continue to exhibit magnetization fluctuations during the observation window after the relaxation period [Fig.~\ref{fig:sector-observables}(c3)]. We examine the origins of these features in turn below.

\subsection{Operator spin-inversion symmetry and field screening}\label{sec:screening}

In the one-dimensional $Q^{(1)}(2,2)$ sector in Fig.~\ref{fig:sector-observables}, the field shifts the energies of the two terminal states without changing the transition operator. By Eq.~\eqref{eq:fix}, the corresponding operator has only two fixed points, the all-up state $\bs_+$ and the all-down state $\bs_-$. For $h\neq0$, one of them is the field-aligned ground state and the other is an antifield high-energy state, with an energy difference of $2|h|N$. Each terminal state is reached by roughly half of the independent trajectories in Fig.~\ref{fig:sector-observables}(a1). The terminal states are uniformly ordered, as shown by $f_{\max}=1$ in Fig.~\ref{fig:sector-observables}(b1). Their magnetization remains constant throughout the observation window, as indicated by $\sigma_M=0$ in Fig.~\ref{fig:sector-observables}(c1). The energy fraction in Fig.~\ref{fig:sector-observables}(d1) distinguishes the field-aligned ground state from the higher-energy antifield state.

The key lies in the symmetry of the operator. Although the external field breaks the spin-inversion symmetry of the energy, the dynamics can retain this symmetry. To express this condition, we define the global spin-inversion operator $\Inv$ by $(\Inv\bs)_i=-\sigma_i$. The single-attempt kernel $P_{h,E_0}$ commutes with $\Inv$ if and only if up and down spins have the same number of open flip channels,
\begin{equation}
P_{h,E_0}\Inv=\Inv P_{h,E_0}
\Longleftrightarrow n_+=n_-.
\label{eq:commute}
\end{equation}
The proof is given in Appendix~\ref{app:commute}. Together with Eq.~\eqref{eq:Q}, this commutation relation implies that the sweep operator satisfies $Q^{(G)}(n,n)\Inv=\Inv Q^{(G)}(n,n)$ when $n_+=n_-=n$. The dynamics thus retains global spin-inversion symmetry even for $h\neq0$, a property we call field screening. Because the diagonal sectors $n_+=n_-$ have finite width in the parameter plane, field screening holds throughout these sectors rather than only on threshold lines.

\begin{figure}[t]
\centering
\includegraphics[width=\columnwidth]{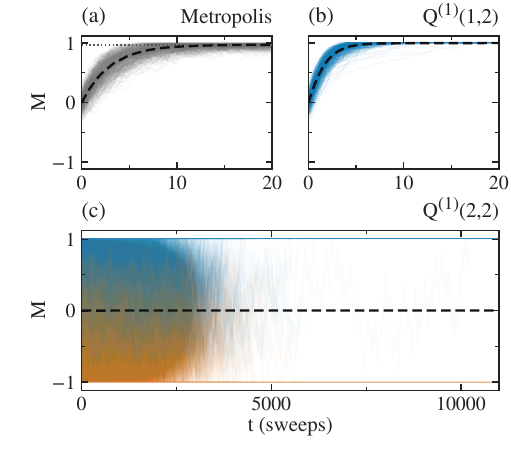}
\caption{Magnetization dynamics with and without field screening. (a)~Under Metropolis dynamics at $T=1$, the mean magnetization approaches the equilibrium value $m_{\rm eq}=0.9679$ (dotted line). (b)~Under $Q^{(1)}(1,2)$ at $E_0=0$, all trajectories reach $M=+1$. (c)~Under $Q^{(1)}(2,2)$ at $E_0=2$, all trajectories reach $M=\pm1$, with nearly equal fractions in the two states. The ensemble mean remains close to zero throughout the relaxation. Thin curves show all trajectories, colored gray in (a) and, in (b,c), blue for those ending at $M=+1$ and orange for those ending at $M=-1$. Black dashed curves show ensemble means. Periodic chains have $L=100$, $J=1$, and $h=0.5$, with unbiased Bernoulli initial states.}
\label{fig:screening}
\end{figure}

For $Q^{(1)}(2,2)$ on a finite one-dimensional periodic chain, this symmetry implies equal absorption probabilities for any initial distribution $\mu$ invariant under spin inversion. To establish this result, we first show that absorption is certain. Flipping a boundary spin moves a domain wall, while flipping an isolated spin annihilates a pair of walls. These are the local rules of zero-field, zero-temperature Metropolis dynamics with random asynchronous updates~\cite{Amar1990,Radicchi2007}. From any nonuniform configuration, a finite sequence of these flips can reach a uniform state with positive probability. The two uniform states are therefore the only closed communicating classes, and the finite state space guarantees eventual absorption~\cite{Norris1997}. Since $\mu$ is invariant under spin inversion, the operator symmetry pairs each trajectory reaching $\bs_+$ with an equally probable inverted trajectory reaching $\bs_-$. The two absorption probabilities, or basin weights under $\mu$, are thus equal and sum to one. Defining $\tau_x$ as the first sweep at which the chain reaches state $x$, we obtain
\begin{equation}
\Pr_{\mu,Q}(\tau_{\bs_+}<\infty)
=
\Pr_{\mu,Q}(\tau_{\bs_-}<\infty)
=\frac12.
\label{eq:half}
\end{equation}
Under the same positive field, finite-temperature Metropolis dynamics relaxes to a positive equilibrium magnetization, as shown in Fig.~\ref{fig:screening}(a). For $Q^{(1)}(2,2)$, however, the $100\,000$ independent trajectories in Fig.~\ref{fig:screening}(c) yield a field-aligned absorption fraction of $0.50035$. The exact 95\% binomial confidence interval, $(0.49725,0.50345)$, is consistent with the equal absorption probabilities in Eq.~\eqref{eq:half}.

Closing only the $(k=1,\sigma=+1)$ channel changes $Q^{(1)}(2,2)$ to $Q^{(1)}(1,2)$, breaking spin-inversion symmetry while preserving the two fixed points. An up spin with one aligned neighbor can no longer flip, so any adjacent up pair becomes permanent. These pairs can grow into larger up domains as neighboring down spins flip up. Absorption remains certain because every nonuniform configuration still has an allowed path to a uniform state. The chain can now reach $\bs_-$ only if no adjacent up pair is present initially or forms during the evolution. For independent unbiased Bernoulli initial states on a periodic chain of length $L\geq3$, let $b_L$ denote this absorption probability. In Appendix~\ref{app:basin}, we derive $b_L$ exactly by accounting for the competition between the disappearance of isolated up spins and the formation of permanent up pairs. For large $L$, we obtain
\begin{equation}
b_L=\left(\frac{2}{\pi}\right)^L
\left[1+\left(-\frac13\right)^L+O(5^{-L})\right].
\label{eq:bL}
\end{equation}
For $h>0$ in this sector, the antifield state remains a fixed point, but its basin weight decays exponentially with $L$. The all-up absorption probability $1-b_L$ therefore approaches one. At $L=100$, $b_L\simeq2.44\times10^{-20}$, consistent with all 512 trajectories reaching the all-up state in Fig.~\ref{fig:screening}(b). Closing a single channel thus replaces an equal split between the two terminal states with almost certain absorption in the field-aligned state, while leaving the fixed points unchanged.

\subsection{Multiple fixed points and frozen morphologies}\label{sec:morphology}

\begin{figure}[t]
\centering
\includegraphics[width=\columnwidth]{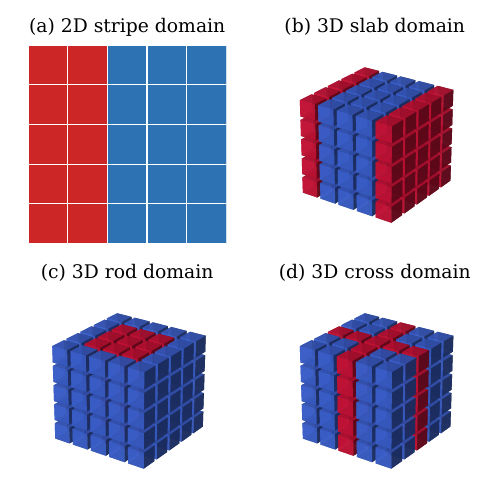}
\caption{Representative nonuniform fixed points and frozen morphologies. Red and blue indicate opposite spin orientations on periodic lattices. (a)~Two-dimensional stripe domain under $Q^{(2)}(3,3)$ with the minimum stable width of 2. (b)--(d)~Three-dimensional frozen morphologies under $Q^{(3)}(4,4)$, showing (b) a planar slab, (c) a square rod, and (d) an intersecting cross. All interface sites have sufficient aligned neighbors ($k\geq n_\sigma$) to satisfy the local stability condition of Eq.~\eqref{eq:fix}.}
\label{fig:frozen}
\end{figure}

The one-dimensional operator $Q^{(1)}(2,2)$ discussed in Sec.~\ref{sec:screening} has only two fixed points, the all-up and all-down states. In two and three dimensions, however, an operator can have many nonuniform fixed points. For example, $Q^{(2)}(3,3)$ admits the stripe state in Fig.~\ref{fig:frozen}(a), while $Q^{(3)}(4,4)$ admits the slab, rod, and cross configurations in Figs.~\ref{fig:frozen}(b)--\ref{fig:frozen}(d). These configurations all satisfy the local stability condition in Eq.~\eqref{eq:fix}, so the system remains frozen once it reaches any of them.

A single operator can admit a large number of fixed points that differ in domain widths and positions. For example, consider $Q^{(2)}(3,3)$ on an $L\times L$ periodic square lattice with $L\geq3$, and let $F_L$ denote its total number of fixed points, counted as distinct spin assignments on labeled lattice sites. At each fixed point, every spin has at most one antialigned neighbor. This constraint rules out interface corners, crossings, and stripes of width 1. Apart from the two uniform states, the fixed points can therefore only be straight stripes in which every domain has width at least 2. For a fixed stripe orientation, up and down domains alternate around the periodic lattice, so there are equally many of each sign. Let $m$ denote this number, giving $2m$ domains in total. Each domain has width at least 2, which requires $4m\leq L$ and hence $1\leq m\leq\lfloor L/4\rfloor$. The widths sum to $L$. Subtracting 2 from each leaves $L-4m$ sites to distribute among $2m$ domains, giving $\binom{L-2m-1}{2m-1}$ ordered width sequences. For each sequence, there are $L$ choices for the starting position and two for the spin of the first domain. This counts each configuration $2m$ times, once for each choice of the first domain, so the number of configurations for this orientation is $\frac{L}{m}\binom{L-2m-1}{2m-1}$. Since nonuniform horizontal and vertical stripes are distinct, summing over $m$ and both orientations and adding the two uniform states gives
\begin{equation}
F_L=\left|\Fix\!\left(Q^{(2)}(3,3)\right)\right|
=2+\sum_{m=1}^{\lfloor L/4\rfloor}
\frac{2L}{m}\binom{L-2m-1}{2m-1}.
\label{eq:FL}
\end{equation}
For $L=32$, this gives $F_{32}=9\,741\,690$ fixed points. Only two are uniform states, while the rest are distinct frozen stripe configurations.

We next ask how often the dynamics reaches these nonuniform fixed points from random initial states. In two dimensions, $Q^{(2)}(3,3)$ accepts a spin flip precisely when the spin has $k\leq2$ aligned neighbors, matching the rule of zero-field, zero-temperature Metropolis dynamics~\cite{Barros2009}. This equivalence allows us to predict the total stripe basin weight from the connection between zero-temperature coarsening and critical percolation. For unbiased Bernoulli initial states, Barros \textit{et al.}~\cite{Barros2009} related the domain statistics after early coarsening to those of critical continuum percolation. Subsequent curvature-driven evolution preserves the wrapping topology, so domains that wrap around the periodic system only horizontally or only vertically evolve into straight stripes in the corresponding direction. The total stripe basin weight therefore equals the sum of the probabilities of these two wrapping events. For critical percolation on a square torus, let $\Pi_{\mathrm h}$ and $\Pi_{\mathrm v}$ denote the probabilities of wrapping only horizontally and only vertically, respectively. Both are known analytically to equal $0.169415435\ldots$, giving the predicted stripe probability in the large-system limit~\cite{Barros2009,Olejarz2012}
\begin{equation}
p_{\mathrm{stripe}}^{\mathrm{theory}}
=\Pi_{\mathrm h}+\Pi_{\mathrm v}
=0.338830870\ldots.
\label{eq:pstripe}
\end{equation}
Our simulations on periodic square lattices with $L=24$, $32$, $48$, $64$, and $128$ yield stripe fractions ranging from $33.18\%$ to $34.48\%$, consistent with this prediction. For a fixed lattice and initial distribution, the same stripe probability also applies at nonzero field within the $Q^{(2)}(3,3)$ sector, since the operator is unchanged. 

\subsection{Neutral channels and long-lived activity}\label{sec:blinker}

Nonuniform fixed points can likewise be reached from random initial states in three dimensions, but the relaxation differs from that in two dimensions. To illustrate, we evolve 512 independent unbiased Bernoulli initial states on a periodic cubic lattice with $L=24$ under $Q^{(3)}(4,4)$ at $h=0.5$ and $E_0=2$. By $10^6$ sweeps, 106 trajectories have frozen in nonuniform configurations and 4 have reached uniform states, while the remaining 402 still have flippable sites. In these 402 trajectories, spins continue to flip throughout the late-time interval from $3\times10^5$ to $10^6$ sweeps, with a mean acceptance rate per site of $0.00430$.

This is the blinker activity previously identified in three-dimensional Ising ferromagnets quenched to zero temperature at zero field~\cite{Spirin2001b,Olejarz2011}. At $h=0$ and $E_0=0$, $Q^{(3)}(4,4)$ implements zero-temperature Metropolis dynamics, accepting all flips with $\Delta H_i\leq0$. On the simple cubic lattice, a single-spin flip changes the energy by $\Delta H_i=4(k_i-3)$. An interface spin with three aligned neighbors ($k_i=3$) can therefore flip at no energy cost. After the flip, it still has three aligned and three antialigned neighbors, so the reverse move also costs no energy. Such flips change the local environments of neighboring spins, shifting the locations of flippable sites along the interface. These rearrangements sustain microscopic activity while the energy remains on a plateau.

\begin{figure}[t]
\centering
\includegraphics[width=\columnwidth]{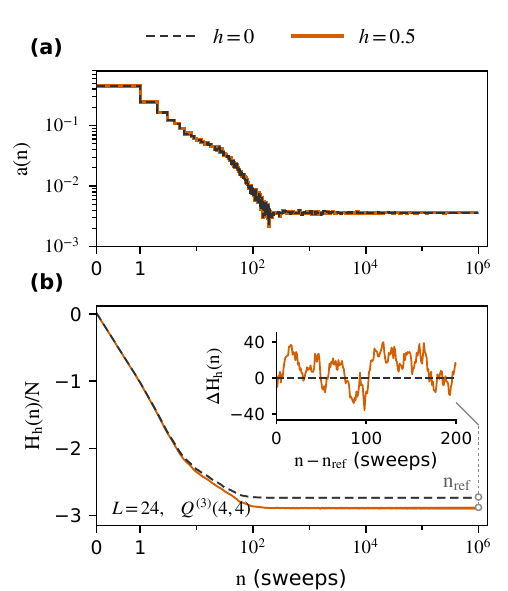}
\caption{Flip activity and energy evolution within the three-dimensional $Q^{(3)}(4,4)$ sector on a periodic cubic lattice ($L=24$). Black dashed and orange solid curves represent $(h,E_0)=(0,0)$ and $(0.5,2)$, sharing a single trajectory with the same initial state and site-selection sequence. (a)~Accepted-flip rate $a(n)$ per attempt, where the two curves coincide window by window, reflecting invariant microscopic dynamics across the sector. (b)~Energy density $H_h(n)/N$ recorded per sweep, with open circles marking $n_{\mathrm{ref}}=10^6$ sweeps. Inset, total energy variation $\Delta H_h(n)$ over the subsequent 200 sweeps, showing a strict energy plateau at zero field versus persistent fluctuations at nonzero field sustained by neutral flip channels.}
\label{fig:blinker}
\end{figure}

The same blinker activity can persist at nonzero field, where it connects configurations of unequal energy. Within the three-dimensional $Q^{(3)}(4,4)$ sector, the operator is unchanged, so the same initial state and site-selection sequence yield identical trajectories at zero and nonzero field. The accepted-flip rates therefore coincide in every observation window and remain nonzero at late times, as shown in Fig.~\ref{fig:blinker}(a). Despite this identical activity, Fig.~\ref{fig:blinker}(b) shows an energy plateau at zero field and continued energy fluctuations at nonzero field. All accepted flips in the 200-sweep window shown in its inset occur in the $k_i=3$ channel and leave the exchange energy unchanged. The total energy change thus comes entirely from the field term, $\Delta H_i=2h\sigma_i$. A flip and its reverse change the energy by opposite amounts but have the same acceptance probability. Neutrality here refers to unchanged exchange energy and equal forward and reverse acceptance probabilities, without requiring conservation of total energy. These properties hold throughout the sector, allowing blinker activity to persist while the total energy fluctuates.

The preceding examples show how a fixed set of open channels governs terminal-state selection, frozen morphologies, and long-lived activity. When the field or threshold crosses a sector boundary, the channels change, and the new operator acts on the configuration produced by the preceding evolution. We next examine how such operator sequences determine which configurations are reached and whether they are held or released as the parameters change.

\section{Ordered Composition of Sector Operators}\label{sec:composition}

\begin{figure*}[t]
\centering
\includegraphics[width=\textwidth]{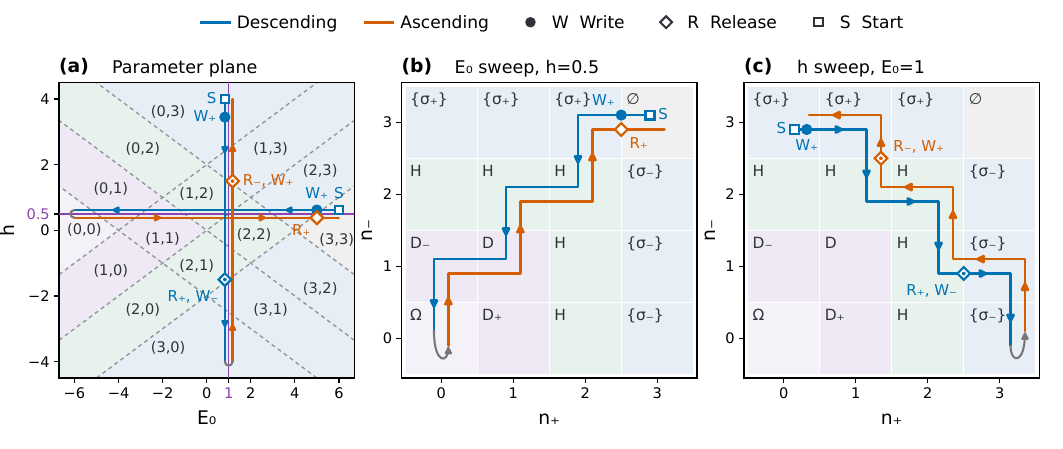}
\caption{Round-trip parameter sweeps and corresponding operator-table paths for a one-dimensional periodic chain ($z=2$). (a)~In the $(E_0,h)$ plane, the six dashed threshold lines of Eq.~\eqref{eq:thresholds} divide the parameter space into 16 sectors labeled by $(n_+,n_-)$. Along the horizontal ($h=0.5$) and vertical ($E_0=1$) purple reference lines, $E_0$ is swept from $6$ to $-6$ and back, and $h$ from $4$ to $-4$ and back. (b)~Corresponding path of the $E_0$ sweep through the $4\times4$ operator table, where the uniform state $\bs_+$ is written and released at the same sector boundary, yielding no hysteresis. (c)~Corresponding path of the $h$ sweep, where $\bs_+$ and $\bs_-$ are written and released at separated boundaries, opening a hysteresis loop. Shading in all panels indicates the fixed-point classes listed in Table~\ref{tab:fix}, with the corresponding fixed-point sets labeled in each cell of (b) and (c). Blue and orange arrows trace the descending and ascending branches from the starting squares ($S$). Filled circles ($W_\pm$) and open diamonds ($R_\pm$) mark where $\bs_\pm$ are written and released, and their superposition in (c) indicates simultaneous release of one state and writing of the other.}
\label{fig:paths}
\end{figure*}

To examine how the dynamics depends on the history of parameter changes, we extend the single-operator evolution of Sec.~\ref{sec:single} to protocols consisting of successive segments, each governed by a fixed sector operator. Starting from an initial distribution $\mu_0$, the distribution after the $m$th segment is
\begin{equation}
\mu_m=\mu_0Q_1^{t_1}\cdots Q_m^{t_m},
\label{eq:protocol}
\end{equation}
where $Q_j=Q^{(G)}(n_{+,j},n_{-,j})$ is the sector operator for segment $j$ on the fixed graph $G$, and the nonnegative integer $t_j$ gives the duration of that segment in sweeps. Because $Q_j$ is unchanged within each sector, the evolution from a given initial distribution depends only on the sequence of sector operators and their dwell times. The same subsequent operator can therefore act on different distributions, depending on the preparation history.

\subsection{Fixed points and round-trip hysteresis}\label{sec:hysteresis}

For a given configuration $x_*$, applying Eq.~\eqref{eq:fix} at every site determines whether $x_*\in\Fix(Q_j)$ in each segment $j$ of the operator sequence. We distinguish four stages along this sequence: (i)~Appearance occurs when $x_*$ first becomes a fixed point of the current operator. (ii)~Writing, however, occurs only when the trajectory reaches $x_*$ during the dwell in a sector where it is a fixed point. (iii)~Holding begins once the state is written and continues as long as each subsequent operator fixes $x_*$. (iv)~Release occurs when the protocol first switches to an operator for which $x_*$ is no longer a fixed point, allowing the system to leave that configuration. Thus, fixed-point membership determines where a state can be written, reachability and dwell time determine whether writing occurs, and the subsequent operator sequence determines how long the written state is held.

\begin{figure}[t]
\centering
\includegraphics[width=\columnwidth]{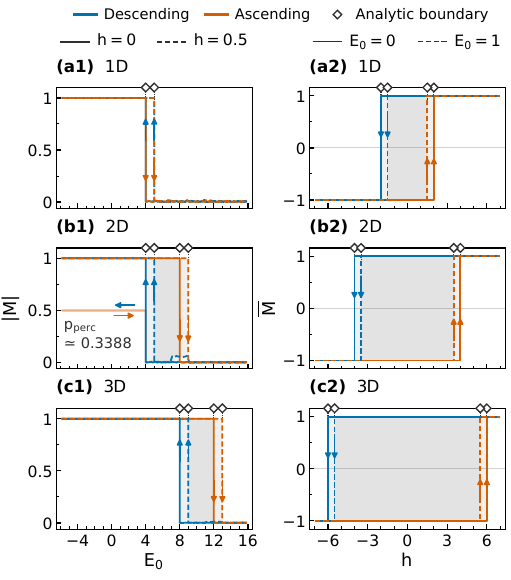}
\caption{Magnetization hysteresis loops under round-trip parameter sweeps across dimensions. Rows correspond to periodic lattices in one to three dimensions with $L=100$, $32$, and $16$. Left column [(a1)--(c1)]: threshold sweeps $E_0$ at fixed field (solid for $h=0$, dashed for $h=0.5$), showing absolute window-averaged magnetization $|\overline M|$. In (b1), the lighter curves show a zero-field trajectory that freezes into a stripe state and exhibits no hysteresis. The annotation $p_{\mathrm{perc}}\simeq0.3388$ gives the direct-quench critical-percolation reference value from Eq.~\eqref{eq:pstripe}. Right column [(a2)--(c2)]: field sweeps $h$ at fixed threshold (solid for $E_0=0$, dashed for $E_0=1$), showing signed magnetization $\overline M$. Blue and orange lines indicate descending and ascending branches, arrows show sweep directions, shading highlights hysteresis areas, and open diamonds mark analytic channel boundaries of Eq.~\eqref{eq:thresholds}. Both sweeps use parameter increments of $0.25$, with the final configuration after sampling at each point carried over to the next. At each point, the magnetization is averaged over 200 sweeps following 4800, 6800, or 8800 relaxation sweeps in one, two, or three dimensions, respectively. Vertical segments connect adjacent sampling nodes at the analytic boundaries.}
\label{fig:loops}
\end{figure}

In this framework, sweeping a continuous parameter back and forth traces a path through the finite operator table. The fixed-point sets can then be read off from the successive entries along this path. Using the one-dimensional periodic chain as an example, Fig.~\ref{fig:paths}(a) shows round-trip sweeps of the threshold $E_0$ and the field $h$ in the parameter plane. Their paths through the operator table are shown in Figs.~\ref{fig:paths}(b) and \ref{fig:paths}(c), respectively. At fixed $h=0.5$, sweeping $E_0$ from $6$ to $-6$ and back traces the path in operator space shown in Fig.~\ref{fig:paths}(b). The path starts at $Q^{(1)}(3,3)$, which has no fixed point. As $E_0$ decreases, the path enters $Q^{(1)}(2,3)$, whose only fixed point is the all-up state $\bs_+$. During the dwell in this sector, the system relaxes to $\bs_+$, completing the writing step marked by $W_+$. The sweep then passes through $Q^{(1)}(2,2)$, $Q^{(1)}(1,2)$, $Q^{(1)}(1,1)$, $Q^{(1)}(0,1)$, and $Q^{(1)}(0,0)$. By Eq.~\eqref{eq:nest}, the fixed-point set can only expand along this sequence, so the system remains in the all-up state. The ascending branch returns along the same path and also leaves the state unchanged. Only upon reentering $Q^{(1)}(3,3)$, which has no fixed point, is the all-up state released, as marked by $R_+$. Over the whole path, the writing and release operators lie at the same operator boundary, so no hysteresis occurs in this case. 

We next fix $E_0=1$ and sweep the field $h$ from $4$ to $-4$ and back, tracing the path shown in Figs.~\ref{fig:paths}(a) and \ref{fig:paths}(c). The path starts at $Q^{(1)}(0,3)$, whose only fixed point is the all-up state $\bs_+$. Starting from a random configuration, the system relaxes to this state. As $h$ decreases, the path passes through $Q^{(1)}(1,3)$, $Q^{(1)}(1,2)$, $Q^{(1)}(2,2)$, and $Q^{(1)}(2,1)$. All four operators retain $\bs_+$ as a fixed point, so the all-up state is held until the path enters $Q^{(1)}(3,1)$, whose only fixed point is the all-down state $\bs_-$. Here the all-up state is released, and the system relaxes to $\bs_-$, completing the release and writing steps marked by $R_+$ and $W_-$. The ascending sweep begins at $Q^{(1)}(3,0)$ and retraces the path through $Q^{(1)}(3,1)$, $Q^{(1)}(2,1)$, $Q^{(1)}(2,2)$, and $Q^{(1)}(1,2)$. The all-down state $\bs_-$ remains a fixed point throughout this sequence, so the system stays in this state until the path enters $Q^{(1)}(1,3)$, where the all-up state is the only fixed point. The all-down state is then released (marked by $R_-$), and the system relaxes to $\bs_+$, writing the all-up state again (marked by $W_+$). Writing and release of each state thus occur at different boundaries in operator space, producing hysteresis.

\begin{figure}[t]
\centering
\includegraphics[width=\columnwidth]{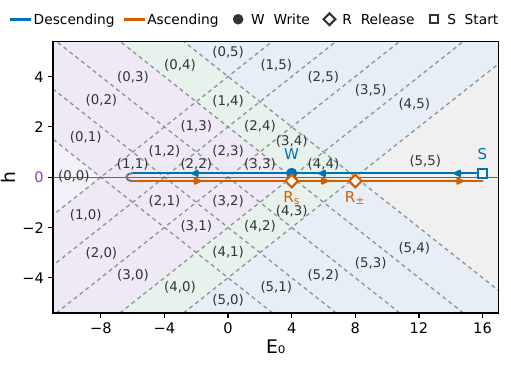}
\caption{Operator sectors and zero-field round-trip sweep in the $(E_0,h)$ plane for the square lattice ($z=4$). Dashed lines from Eq.~\eqref{eq:thresholds} divide the parameter plane into sectors labeled by $(n_+,n_-)$. Shading indicates sectors with no fixed points (gray), only uniform fixed points (blue and green), both uniform and nonuniform fixed points (purple), or all configurations fixed (pale purple). Blue and orange arrows trace the descending and ascending branches along $h=0$, from $S$ at $E_0=16$ to $-6$ and back, with a vertical offset for clarity. Crossing $E_0=4$ at $W$ brings the system into $Q^{(2)}(3,3)$, where writing occurs during the subsequent dwell. On the return sweep, stripe states are released at the same boundary ($R_{\rm s}$ at $E_0=4$), leaving no holding interval in $Q^{(2)}(4,4)$ and hence no hysteresis. By contrast, ordered states $\bs_\pm$ remain fixed throughout this sector and are released only at $E_0=8$ ($R_\pm$), producing the hysteresis loop in Fig.~\ref{fig:loops}(b1).}
\label{fig:2d-plane}
\end{figure}

Figure~\ref{fig:loops} shows the resulting magnetization curves in one, two, and three dimensions. Consistent with the mechanism described above, the one-dimensional system shows no hysteresis when $E_0$ is swept at fixed $h$, as shown in Fig.~\ref{fig:loops}(a1). The location of this jump shifts with $h$ according to Eq.~\eqref{eq:thresholds}. When $h$ is swept at fixed $E_0$, writing and release instead occur at different sector boundaries, producing the hysteresis loops shown in Fig.~\ref{fig:loops}(a2). The hysteresis shown in Fig.~\ref{fig:loops}(b1)--(c2) for two- and three-dimensional systems arises from the same mechanism. Note, however, that the fixed-point criterion only identifies the sectors in which a given configuration is absorbing. Whether and when that configuration is written depends on the preparation, system size, and dwell times. This distinction is particularly evident in Fig.~\ref{fig:loops}(b1), where the occurrence of hysteresis is probabilistic even under identical sweep conditions. Figure~\ref{fig:2d-plane} shows how the same path in the $(E_0,h)$ plane can lead to either response. The blue and orange arrows mark the descending and ascending sweeps of $E_0$ at $h=0$, respectively. Starting from $S$, the descending sweep passes successively through $Q^{(2)}(5,5)$, $Q^{(2)}(4,4)$, and $Q^{(2)}(3,3)$. Although the fully ordered states $\bs_\pm$ are already fixed points in $Q^{(2)}(4,4)$, the system is unlikely to reach either state within the dwell time. Even spins aligned with three of their four neighbors can still flip here, sustaining interfacial fluctuations that hinder complete ordering. Once the sweep crosses the $E_0=4$ boundary marked by $W$ and enters $Q^{(2)}(3,3)$, these flips are blocked, leaving only flips that lower or conserve the energy. Domains then coarsen, and the $L=32$ system shown in Fig.~\ref{fig:loops}(b1) freezes into a fully ordered state or a stripe state within the dwell time, completing the writing step. Once written, both types of state remain fixed for the rest of the descending sweep, but they respond differently on the return sweep. A fully ordered state remains fixed throughout $Q^{(2)}(4,4)$ and is released only upon entering $Q^{(2)}(5,5)$, which has no fixed point. This release occurs at the $E_0=8$ boundary, marked by $R_\pm$ in Fig.~\ref{fig:2d-plane}. The interval between $W$ and $R_\pm$ therefore spans $Q^{(2)}(4,4)$, where the ordered state is held on the ascending branch, producing hysteresis. A stripe state, however, is released as soon as the ascending sweep reenters $Q^{(2)}(4,4)$, where stripes are no longer fixed points. Its release marker $R_{\rm s}$ lies at the same $E_0=4$ boundary as $W$, so there is no holding interval across $Q^{(2)}(4,4)$. Thus, when the system freezes into a stripe state, it shows no such hysteresis. Under the zero-field protocol of Fig.~\ref{fig:loops}, independent ensembles of $100\,000$ trajectories each at $L=24,32,48,64,128$ yield stripe fractions of 33.18\%, 33.23\%, 33.65\%, 33.59\%, and 33.94\%, respectively. The stripe fraction thus remains close to $1/3$ across the sizes studied. Small deviations from the direct-quench prediction in Eq.~\eqref{eq:pstripe} may reflect finite-size effects and correlations generated during the preceding evolution under $Q^{(2)}(4,4)$.

\subsection{Blocked domain interiors and Mpemba-like relaxation}\label{sec:mpemba}

In the hysteresis protocols of Sec.~\ref{sec:hysteresis}, a prepared configuration remains frozen as long as each subsequent operator leaves it unchanged. When a new operator no longer has that configuration as a fixed point, relaxation resumes. Here we consider a different problem, in which several prepared initial states evolve under the same target operator. If this target operator drives all of them to a common fixed point with probability one, how do their relaxation dynamics differ? In particular, can a state initially farther from this fixed point nevertheless reach it sooner on average than one that starts closer? These questions capture the central idea of the Mpemba effect~\cite{Mpemba1969}.

\begin{figure}[t]
\centering
\includegraphics[width=\columnwidth]{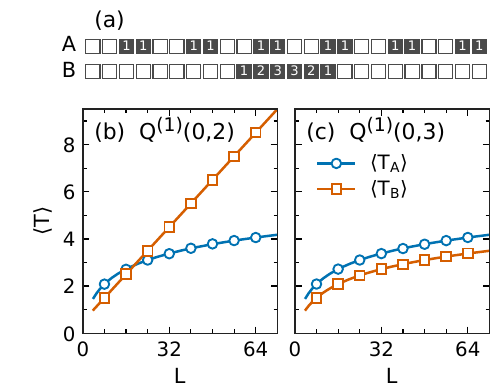}
\caption{Mean absorption times of two preparations on a periodic chain. (a)~Configurations A and B for $L=24$, with open and filled squares denoting up and down spins. The numbers indicate the earliest stage at which each down spin can flip under $Q^{(1)}(0,2)$. (b),(c)~Mean absorption times $\langle T_A\rangle$ and $\langle T_B\rangle$ versus chain length $L$ under $Q^{(1)}(0,2)$ and $Q^{(1)}(0,3)$, respectively. Lines show Eqs.~\eqref{eq:exact1d} and \eqref{eq:exact1d-bulk}, and symbols show simulations with $5\times10^4$ trajectories.}
\label{fig:chain-ab}
\end{figure}

On a periodic chain, a configuration with more down spins can reach the all-up state $\bs_+$ sooner on average. To illustrate this, we compare two initial configurations on a chain of length $L\geq8$, with $L$ a multiple of 4. Configuration A consists of repeated $++--$ blocks, whereas B contains a single down domain of length $L/4$ surrounded by up spins, as shown in Fig.~\ref{fig:chain-ab}(a). Their misaligned fractions $d_+=(1-M)/2$ are $1/2$ and $1/4$, respectively, so A starts twice as far from $\bs_+$ by this measure. Neither configuration contains an isolated spin, so both are fixed points of $Q^{(1)}(1,1)$. We evolve both configurations under $Q^{(1)}(0,2)$, for which a selected down spin flips if it has at least one up neighbor, while up spins remain fixed. Both configurations therefore reach $\bs_+$ with probability one. To compare their relaxation, we define the absorption time $T$ as the number of update attempts required to first reach $\bs_+$, divided by $N$. The mean absorption times, measured in sweeps and derived in Appendix~\ref{app:release}, are
\begin{equation}
\langle T_A\rangle_{Q^{(1)}(0,2)}=H_{L/2},
\qquad
\langle T_B\rangle_{Q^{(1)}(0,2)}=\frac{L}{8}+\frac12,
\label{eq:exact1d}
\end{equation}
where $H_m=\sum_{j=1}^{m}j^{-1}$ is the $m$th harmonic number. The mean absorption time grows as $\ln L$ for A but linearly with $L$ for B, as shown in Fig.~\ref{fig:chain-ab}(b). A therefore reaches $\bs_+$ sooner on average than B for sufficiently large $L$, even though it starts with twice as many down spins.

The difference lies in how the down domains disappear. In A, every down spin already has an up neighbor and flips permanently when first selected. Reaching the absorbing state therefore requires only that each of the $L/2$ down sites be selected at least once. In B, by contrast, only the spins at the two ends of the down domain can initially flip. Each flip moves a boundary inward, allowing the next spin to flip. Thus, each down spin in A can flip without waiting for another spin to flip, whereas spins inside the down domain in B must wait for a boundary to reach them. Opening the bulk channel $(k=2,\sigma=-1)$ removes this delay by allowing a down spin to flip even when both neighbors are down. Under the resulting target operator $Q^{(1)}(0,3)$, every down spin in either preparation flips up on its first selection and remains up. The mean absorption times therefore depend only on the initial number of down spins,
\begin{equation}
\langle T_A\rangle_{Q^{(1)}(0,3)}=H_{L/2},
\qquad
\langle T_B\rangle_{Q^{(1)}(0,3)}=H_{L/4}.
\label{eq:exact1d-bulk}
\end{equation}
B now reaches the absorbing state sooner on average for every allowed $L$, as shown in Fig.~\ref{fig:chain-ab}(c). The target operators $Q^{(1)}(0,2)$ and $Q^{(1)}(0,3)$ can be realized at $(h,E_0)=(3,-4)$ and $(3,0)$, respectively. Thus, opening a single channel reverses the ordering of the mean absorption times for sufficiently large $L$.

\begin{figure}[!t]
\centering
\includegraphics[width=\columnwidth]{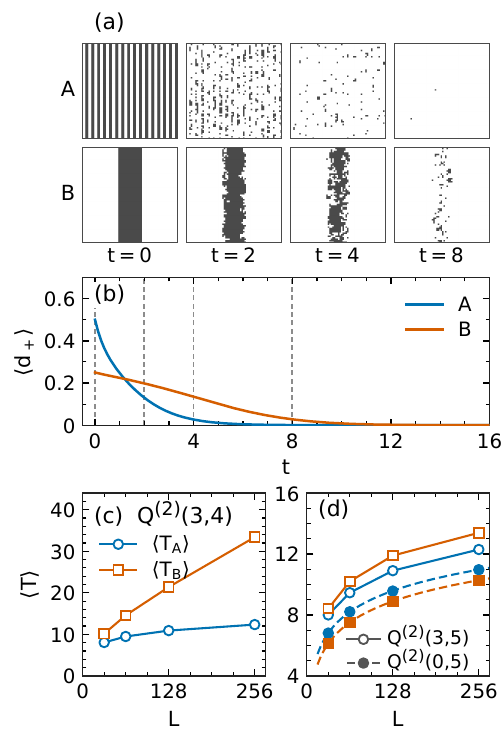}
\caption{Relaxation dynamics and kinetic bottleneck of stripe preparations on a periodic square lattice.
(a)~Snapshots of A (alternating width-2 stripes) and B (a single down stripe of width $L/4$) under $Q^{(2)}(3,4)$ at $L=64$ and $t=0$, 2, 4, and 8 sweeps. White and gray denote up and down spins. The thin stripes in A break up across their width, whereas the thick stripe in B erodes inward.
(b)~Mean misaligned fraction $\langle d_+\rangle$ versus time under the same conditions. Although A starts with twice the down-spin fraction of B, the curves cross at $t\approx1.2$ sweeps. Vertical dashed lines mark the snapshot times.
(c)~Mean absorption times $\langle T_A\rangle$ (circles) and $\langle T_B\rangle$ (squares) versus $L$ under $Q^{(2)}(3,4)$. A reaches the all-up state sooner on average, and the gap grows with $L$.
(d)~Mean absorption times versus $L$ under $Q^{(2)}(3,5)$ (open symbols) and $Q^{(2)}(0,5)$ (filled symbols), with circles and squares denoting A and B. Dashed curves show the exact means $H_{L^2/2}$ and $H_{L^2/4}$ under $Q^{(2)}(0,5)$. Opening the bulk channel reduces the gap, and suppressing back flips restores the ordering $\langle T_B\rangle < \langle T_A\rangle$.}
\label{fig:stripes}
\end{figure}

The delay in B reflects a local constraint that applies on any $z$-regular graph. A down spin with $u$ up neighbors has $k=z-u$ aligned neighbors, so the local condition in Eq.~\eqref{eq:fix} permits it to flip precisely when
\begin{equation}
u\geq\theta_-=z-n_-+1.
\label{eq:theta}
\end{equation}
When $\theta_-\geq1$, a spin in the interior of a down domain must wait until enough of its neighbors have flipped up. Starting from the initially up sites, we can repeatedly add sites with at least $\theta_-$ neighbors already in the set. The resulting set is the bootstrap-percolation closure~\cite{Chalupa1979,Janson2012}, a construction also used to describe kinetic constraints in facilitated spin models~\cite{Sellitto2005}. It constrains the order in which initially down spins can first flip up, even when up spins can later flip back down. In particular, the mean time for an initially down spin to first flip up is bounded below by a constant times its distance from the initial up region (Appendix~\ref{app:release}). Neither the misaligned fraction nor the energy captures the dependencies between flips that give rise to this bound. Opening the bulk channel $(k=z,\sigma=-1)$ sets $\theta_-=0$, allowing every initially down spin to flip on its first selection without waiting for its neighbors. This also allows spins to flip in the all-down state, which is therefore no longer a fixed point.

The reversal in mean absorption times also occurs on a periodic square lattice, even when up spins can flip back down. For $L$ a multiple of 8, we take A to consist of alternating up and down stripes of width 2, and B to contain a single down stripe of width $L/4$. Both preparations are fixed points of $Q^{(2)}(3,3)$, with initial misaligned fractions $1/2$ and $1/4$, respectively. We then evolve both under $Q^{(2)}(3,4)$, which can be realized at $(h,E_0)=(0.5,4)$ and has $\theta_-=1$. Under this operator, every spin in an initial up stripe remains up because it has at least three neighbors within that stripe. Together with $\theta_-=1$, this guarantees absorption in $\bs_+$ for both preparations (Appendix~\ref{app:release}). As shown in Fig.~\ref{fig:stripes}(a), the thin down stripes in A break up across their full width, whereas the stripe in B shrinks inward from its edges. Figure~\ref{fig:stripes}(b) shows that their mean misaligned fractions cross during relaxation at $L=64$. Figure~\ref{fig:stripes}(c) shows that $\langle T_B\rangle$ increases much faster with $L$ than $\langle T_A\rangle$. For example, at $L=128$, the mean absorption times are $10.90\pm0.03$ sweeps for A and $21.38\pm0.03$ sweeps for B, with uncertainties denoting standard errors. A therefore again reaches $\bs_+$ sooner on average despite starting with twice as many down spins. The delay of B grows from $2.12$ sweeps at $L=32$ to $21.05$ sweeps at $L=256$.

Opening the bulk channel at $(h,E_0)=(1.5,6)$ gives $Q^{(2)}(3,5)$. In A, every initially down spin has a neighbor in an initial up stripe that remains fixed, so it never needs this channel to flip. A therefore evolves exactly as under $Q^{(2)}(3,4)$, while the additional channel accelerates B. At $L=128$, the delay of B relative to A falls to $0.99\pm0.04$ sweeps, as shown in Fig.~\ref{fig:stripes}(d), although newly flipped up spins can still flip back down. Suppressing these reverse flips at $(h,E_0)=(5,0)$ gives $Q^{(2)}(0,5)$, under which every down spin flips up permanently when first selected. The mean absorption time is then exactly $H_m$, independent of how the $m$ initial down spins are arranged (Appendix~\ref{app:release}). For the two stripe preparations, this gives
\begin{equation}
\langle T_A\rangle_{Q^{(2)}(0,5)}=H_{L^2/2},
\qquad
\langle T_B\rangle_{Q^{(2)}(0,5)}=H_{L^2/4}.
\end{equation}
B therefore reaches $\bs_+$ sooner on average because it starts with fewer down spins. Figure~\ref{fig:stripes}(d) compares these exact results (dashed lines) with the simulations (filled symbols).

\begin{figure}[!t]
\centering
\includegraphics[width=\columnwidth]{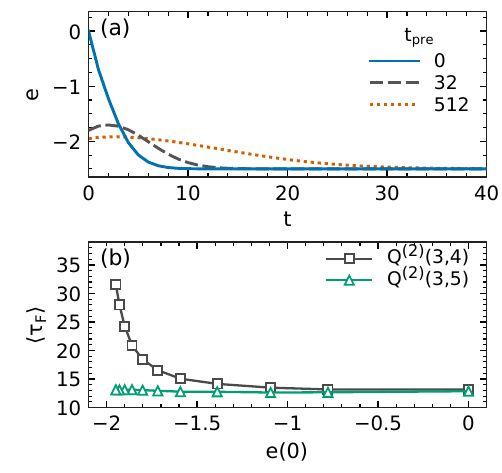}
\caption{Anomalous relaxation after coarsening and its suppression by bulk
flips on a square lattice ($L=128$).
(a)~Mean energy density $e$ versus time $t$ under $Q^{(2)}(3,4)$,
after coarsening under $Q^{(2)}(3,3)$ for $t_{\rm pre}=0$ (solid),
32 (dashed), and 512 sweeps (dotted). Energies are evaluated at
$h=0.5$. The highest-energy curve crosses the others at
$t\approx3$--$4$ sweeps, and this preparation freezes soonest on average.
(b)~Mean freezing time $\langle\tau_{\rm F}\rangle$ versus initial
energy density $e(0)$ under $Q^{(2)}(3,4)$ (squares) and
$Q^{(2)}(3,5)$ (triangles). Lower-energy preparations freeze later
under constrained dynamics, whereas opening the bulk channel yields
nearly constant mean freezing times.
Each preparation uses 1000 trajectories.}
\label{fig:coarsening}
\end{figure}

A related reversal occurs for states prepared by natural coarsening. Starting from unbiased Bernoulli initial states, we evolve the system under $Q^{(2)}(3,3)$ for $t_{\rm pre}$ sweeps before switching to $Q^{(2)}(3,4)$. Longer coarsening produces thicker domains and lowers the energy, while spin-inversion symmetry keeps the mean initial misaligned fraction at $1/2$. We therefore compare the preparations by their initial energies. With the bulk channel closed, the all-down state is also fixed. Unlike the stripe preparations, these states can freeze into either uniform state, so we define the freezing time $\tau_{\rm F}$ as the first integer sweep at which the system is in an absorbing state of the target operator. The all-down outcome accounts for at most $0.1\%$ of trajectories for each preparation. Figure~\ref{fig:coarsening}(a) shows the energy curves at $L=128$ for $t_{\rm pre}=0$, $32$, and $512$ sweeps. The corresponding initial energy densities are approximately $0$, $-1.80$, and $-1.95$, and the mean freezing times are $13.2$, $18.5$, and $31.6$ sweeps, respectively. The preparation with the highest initial energy thus freezes soonest on average, and its energy curve crosses the other two between the third and fourth sweeps. Figure~\ref{fig:coarsening}(b) shows that the mean freezing time increases steadily as the initial energy decreases. Opening the bulk channel, which gives $Q^{(2)}(3,5)$, largely removes this dependence, yielding mean freezing times of $12.6$--$13.1$ sweeps across all preparation times studied.

For the preparation families considered here, system size determines both whether the relaxation anomaly occurs and how pronounced it is. Figure~\ref{fig:chain-ab}(b) shows that B, which starts closer to $\bs_+$, reaches it sooner on average for the allowed chain lengths $L\leq16$, whereas A arrives sooner for $L\geq20$. Increasing $L$ thus reverses the relaxation ordering without changing the local dynamics. Size-dependent Mpemba regions have also been reported in the mean-field Ising antiferromagnet, where both the boundaries between relaxation regimes and the initial temperature at which the strong Mpemba effect occurs shift with system size~\cite{Klich2019}. In the present model, this size dependence follows directly from the geometry of the initial states. As $L$ increases, the blocked domain in B becomes wider, while the domains in A remain two sites wide. Equation~\eqref{eq:exact1d} therefore gives a mean absorption time that grows logarithmically with $L$ for A but linearly for B. On the square lattice, the gap between the mean absorption times grows with $L$, as shown in Fig.~\ref{fig:stripes}(c), indicating that increasing system size also strengthens the anomaly.

The anomaly can also be controlled by varying the parameters to open or close flip channels. A similar dependence on parameters occurs in the mean-field antiferromagnet, whose temperature--field diagram contains regions with and without Mpemba effects~\cite{Klich2019}. In reaction networks, varying transition rates can likewise create or eliminate regions where the strong Mpemba effect occurs~\cite{Bera2026}. In the present model, switching between such regimes has a direct local origin because each sector boundary opens or closes a specific flip channel. On the periodic chain, moving from $Q^{(1)}(0,2)$ to $Q^{(1)}(0,3)$ opens the bulk channel and restores normal ordering for every allowed $L$, as shown in Fig.~\ref{fig:chain-ab}(c). On the square lattice, Fig.~\ref{fig:stripes}(d) shows that opening this channel in $Q^{(2)}(3,5)$ greatly reduces the anomaly, while additionally suppressing reverse flips in $Q^{(2)}(0,5)$ reverses the ordering of A and B. Because the operator is constant within each $(h,E_0)$ sector, the full absorption statistics remain unchanged throughout that sector for fixed preparations and system size. The relaxation orderings found above therefore hold over entire parameter regions, and changes in ordering can be traced to the opening or closing of specific flip channels.

The effects of system size and channel opening link the relaxation anomaly to the same local rule that governs hysteresis. This neighbor-count rule determines both whether a prepared state remains frozen and which spins must wait for their neighbors to flip after release. By identifying the spins responsible for the delay, it explains how domain geometry amplifies the delay and how opening a flip channel removes it. This local description complements earlier accounts based on slow modes and mean first-passage times~\cite{Lu2017,Klich2019,Walker2022} and on initial spatial correlations in spin systems~\cite{Vadakkayil2021,Chatterjee2024}. For both preparations in the chain, the slowest step is the removal of the last down spin, which flips with probability $1/L$ per update attempt. Their survival probabilities therefore share the same asymptotic decay factor $(1-1/L)^k$, where $k$ counts update attempts. The reversal in mean absorption times thus occurs without eliminating the slowest relaxation mode.

\section{Conclusion}\label{sec:conclusion}

In this work, we use the $\delta$-Ising model to show that combinations of flip channels beyond those available at zero temperature have physical consequences of their own. Most directly, an energetic preference need not produce a kinetic bias. When up and down spins have the same open channels, the transition operator commutes with global spin inversion even at nonzero field. Under this field screening, stripe freezing in two dimensions and blinker activity in three dimensions persist in a field. In the symmetric coarsening sector of a periodic chain, unbiased random initial states reach the field-aligned and antifield uniform states with probability $1/2$ each, despite their energy difference. Closing a single channel leaves both states fixed, yet the antifield basin weight under the same initial ensemble decays as $(2/\pi)^L$ with chain length $L$. Thus, closing a single channel at fixed field raises the probability of reaching the ground state from $1/2$ to nearly one for large $L$.

Ordered products of sector operators separate reaching a state from holding it. A state reached in one sector can remain fixed in later sectors where random initial states would rarely reach it. Round-trip sweeps therefore exhibit hysteresis when the system reaches a state at one sector boundary and releases it at another. Even along the same sweep path, the response can depend on which state is reached. During zero-field threshold sweeps on a square lattice with finite dwell times, the system reaches a uniform state and exhibits hysteresis with probability about $2/3$. With probability about $1/3$, it instead freezes into a stripe state and shows no hysteresis because this state is released on the return sweep at the same sector boundary at which it formed. When prepared states are instead switched into a common target operator, its open channels decide which flips must wait for others. If down spins can flip at domain boundaries but not inside domains, interior spins must wait for the boundaries to reach them. Thus, with the same all-up final state, many thin down-spin domains can disappear sooner on average than a single thick domain containing fewer down spins. On a periodic chain, their mean absorption times grow as $\ln L$ and $L$, respectively, so a Mpemba-like reversal sets in as the system grows. Opening the bulk flip channel, however, eliminates this reversal on the chain and largely suppresses it on the square lattice.

Different sectors of the $\delta$-Ising model reproduce the local flip rules of several established stochastic models. For example, freezing down spins gives asynchronous bootstrap percolation, with $Q^{(2)}(3,0)$ implementing 3-core pruning of the up-spin set~\cite{Chalupa1979}. At threshold one, $Q^{(G)}(z,0)$ grows the down-spin set by adding each perimeter site at unit rate, giving continuous-time Eden growth (version A)~\cite{Manin2023}. Allowing down spins to flip regardless of their neighbors, as in $Q^{(2)}(3,5)$, instead gives a threshold contact process with equal infection and recovery rates~\cite{CoxDurrett1991,FontesSchonmann2008}. The diagonal sectors $Q^{(G)}(n,n)$ with $1\leq n\leq z$ correspond to threshold voter models~\cite{DurrettSteif1993,Liggett1994}. On the chain, $Q^{(1)}(2,2)$ is also the vacillating voter model, in which a spin flips if either neighbor disagrees~\cite{LambiotteRedner2007}. On the square lattice, $Q^{(2)}(4,4)$ is the threshold-one voter model, which supports coexistence of both spin orientations on the infinite lattice~\cite{Liggett1994}. This coexistence helps explain why ordered states are difficult to write in this sector. In sectors on the three antidiagonals $n_++n_-=z$, $z+1$, and $z+2$, the local flip rates match those of zero-temperature Ising dynamics in an effective field $h'=n_--n_+$, with zero-energy flips rejected, absent, and accepted at unit rate, respectively~\cite{GodrecheLuck2005}. On the chain, the first case with $h'=0$ is $Q^{(1)}(1,1)$, which maps in bond variables to random sequential adsorption of dimers~\cite{GodrecheLuck2005,Krapivsky2010}. For $E_0<0$, the acceptance criterion also matches the switching rule for interacting hysterons with uniform coercivity, whose transition graphs describe their hysteretic response~\cite{Teunisse2025,MunganTerzi2019}. The $\delta$-Ising model thus embeds these stochastic dynamics and hysteron switching rules in a common operator table, with varying $h$ and $E_0$ selecting the dynamics.

Finally, our results suggest several directions for further work. First, although our lower bound on mean first-flip times grows linearly with distance, a quantitative theory of absorption times in two and three dimensions remains to be developed. Second, softening the hard threshold with noise or thermal fluctuations would clarify how these perturbations affect the parameter ranges over which the observed phenomena persist. Third, studying the $\delta$-Ising model in higher dimensions would test whether qualitatively new forms of collective behavior emerge.

\begin{acknowledgments}
This work was supported by the National Natural Science Foundation of China (12675325, 12675259, and W2621001).
\end{acknowledgments}

\section*{Data Availability}

The data and reproduction code that support the findings of this article are not publicly available. They are available from the authors upon reasonable request.

\appendix

\section{Proof of the spin-inversion commutation criterion}\label{app:commute}

We use the notation of Sec.~\ref{sec:model} and write $P\equiv P_{h,E_0}$. Since $\Inv$ is a permutation of $\Omega_G$, $(P\Inv)(\bs,\bs')=P(\bs,\Inv\bs')$ and $(\Inv P)(\bs,\bs')=P(\Inv\bs,\bs')$, so $P\Inv=\Inv P$ is equivalent to
\begin{equation}
P(\bs,\bs')=P(\Inv\bs,\Inv\bs'),\qquad \forall\,\bs,\bs'\in\Omega_G.
\label{eq:A1}
\end{equation}

Inversion commutes with single-site flips, $\Inv\bs^{(i)}=(\Inv\bs)^{(i)}$. It also reverses vertex $i$ together with its neighbors, so $k_i(\Inv\bs)=k_i(\bs)$ while $\sigma_i$ changes sign. Equation~\eqref{eq:kernel} then gives
\begin{equation}
\begin{aligned}
P(\bs,\bs^{(i)})&=\frac{1}{N}\Theta\!\left(E_0-\Delta H(k_i,\sigma_i)\right),\\
P(\Inv\bs,\Inv\bs^{(i)})&=\frac{1}{N}\Theta\!\left(E_0-\Delta H(k_i,-\sigma_i)\right).
\end{aligned}
\label{eq:A2}
\end{equation}
Elements between configurations that differ at more than one site vanish on both sides of Eq.~\eqref{eq:A1}, and once the flip elements agree, row normalization makes the self-loop elements agree as well. Equation~\eqref{eq:A1} therefore reduces to equality of the two lines of Eq.~\eqref{eq:A2}. Since spins can be assigned freely, every channel $(k,\sigma)$ occurs at some vertex of some configuration, so this equality holds for all $\bs$ and $i$ if and only if each pair of field-split channels is open or closed together,
\begin{equation}
\begin{gathered}
\Theta\!\left(E_0-\Delta H(k,+1)\right)=\Theta\!\left(E_0-\Delta H(k,-1)\right),\\
k=0,1,\ldots,z.
\end{gathered}
\label{eq:A3}
\end{equation}

By Sec.~\ref{sec:sectors}, the open channels of orientation $\sigma$ are exactly those with $k<n_\sigma$. If $n_+=n_-$, Eq.~\eqref{eq:A3} holds. If $n_+\neq n_-$, say $n_+>n_-$, then $n_-\leq z$, and the channel $(n_-,+1)$ is open while $(n_-,-1)$ is closed, so Eq.~\eqref{eq:A3} fails at $k=n_-$. This proves Eq.~\eqref{eq:commute}.

\section{Exact derivation of the one-dimensional antifield basin weight}\label{app:basin}

Consider a periodic chain of length $L\geq3$ and the operator $Q^{(1)}(1,2)$. Let $S$ be the set of positions of up spins, and let $u(S)$ be the probability of eventually reaching the all-down state from this configuration. If $S$ contains adjacent vertices, this pair of up spins can never flip, so $u(S)=0$. The remaining sets $S$ are independent sets of the cycle graph. Let $\partial S$ be the set of down sites adjacent to $S$. There are two kinds of state changes. Either an isolated up spin in $S$ disappears, or a down spin in $\partial S$ flips up and immediately forms an irremovable up pair. Every site has the same proposal probability, and after removing the self-loops produced by rejected flips the first-step equations are
\begin{equation}
\begin{gathered}
u(\varnothing)=1,\qquad
u(S)=\frac{\sum_{i\in S}u(S\smallsetminus\{i\})}{|S|+|\partial S|},\\
b_L=2^{-L}\sum_{S\text{ independent}}u(S).
\end{gathered}
\label{eq:B1}
\end{equation}

To obtain a closed form, we first exclude the fully alternating even rings, which are counted separately below, and divide the remaining independent sets into maximal alternating blocks of the form $+-+\cdots-+$, with adjacent blocks separated by at least two consecutive down spins. Representing random asynchronous updates by independent unit-rate Poisson clocks on the sites gives an embedded attempt chain identical to the original rule and therefore does not change absorption probabilities. Different blocks and their adjacent down spins use nonoverlapping clocks, so their success probabilities multiply. Let $a_m$ be the success probability of a single block containing $m$ up spins, with $a_0=1$. A block has $m$ up spins that can disappear and $m+1$ down spins whose flipping up causes failure. Removing the $(r+1)$th up spin leaves two blocks containing $r$ and $m-1-r$ up spins, so
\begin{equation}
(2m+1)a_m=\sum_{r=0}^{m-1}a_ra_{m-1-r}.
\label{eq:B2}
\end{equation}

Let $A(u)=\sum_{m\geq0}a_mu^m$. Equation~\eqref{eq:B2} gives $2uA'(u)+A(u)-1=uA(u)^2$, and with $A(0)=1$ the unique formal power-series solution is $A(u)=\tan\sqrt u/\sqrt u$.

We now sum over chain lengths and assign the Bernoulli weight $1/2$ to each initial spin. With $x$ marking the number of sites, an alternating block and a gap containing at least two down spins have the generating functions
\begin{equation}
C(x)=\frac{2}{x}\left[A(x^2/4)-1\right],
\qquad G(x)=\frac{x^2}{4(1-x/2)}.
\label{eq:B3}
\end{equation}
Periodic configurations with $j$ blocks are counted by $(CG)^j/j$, and the operator $x\,\mathrm d/\mathrm dx$ then restores the count with fixed site labels. The all-down configuration is counted separately. We must also add the fully alternating even rings, which contain no double-down gap. When $L=2m\geq4$, the first state change removes an up spin with probability $1/2$ and leaves a single block with success probability $a_{m-1}$. The two fully alternating configurations therefore contribute $a_{m-1}(x/2)^{2m}$ in total. Adding the three classes gives a generating function whose coefficients for $L\geq3$ are $b_L$,
\begin{equation}
\begin{aligned}
\mathcal B(x)
&=\frac{1}{1-x/2}
+x\frac{\mathrm d}{\mathrm dx}\left[-\log(1-C(x)G(x))\right]\\
&\quad+\frac{x^2}{4}\left[A(x^2/4)-1\right]\\
&=1-\frac{x^2}{4}+\frac{x}{2}\tan\left(\frac{\pi}{4}+\frac{x}{2}\right).
\end{aligned}
\label{eq:B4}
\end{equation}
Equivalently,
\begin{equation}
b_L=\frac{A_{L-1}}{2(L-1)!},\qquad L\geq3,
\label{eq:bL-euler}
\end{equation}
where $A_n$ are the Euler zigzag numbers, defined by
$\sec x+\tan x=\sum_{n\geq0}A_nx^n/n!$.
The poles of the tangent function lie at $x_k=\pi(4k+1)/2$. Extracting the coefficients of Eq.~\eqref{eq:B4} for $L\geq3$ from its partial-fraction expansion gives the absolutely convergent sum
\begin{equation}
b_L=\sum_{k\in\mathbb Z}x_k^{-L}
=\left(\frac{2}{\pi}\right)^L\sum_{k\in\mathbb Z}(4k+1)^{-L}.
\label{eq:B5}
\end{equation}
Separating the $k=0$ and $k=-1$ terms leaves a remainder of order $O(5^{-L})$, which gives Eq.~\eqref{eq:bL}. The first-step recursion also gives $b_3=1/4$, $b_4=1/6$, $b_5=5/48$, and $b_6=1/15$, in agreement with the closed form.

\section{Common absorption, first-flip bound, and exact absorption times}\label{app:release}

Let $G$ be a finite connected $z$-regular graph. Suppose that the initial up set contains a nonempty subset $C$ in which every $i\in C$ has at least $n_+$ neighbors belonging to $C$. Consider the first down flip in $C$, if any. Before it, all of $C$ is still up, so the flipping site has at least $n_+$ aligned neighbors and cannot flip, a contradiction. Hence $C$ stays up forever. For $n_-=z$ or $z+1$, every mixed configuration that contains $C$ has an allowed path to the all-up state, obtained by successively flipping up down sites on the boundary between up and down regions. In a finite state space in which every reachable state can reach the all-up state, no other closed recurrent class can exist, so the all-up state is reached with probability one and the mean time is finite. The initial up stripes of Sec.~\ref{sec:mpemba} have width at least 2, so each of their sites has at least three neighbors in the same stripe, which satisfies this condition for $Q^{(2)}(3,4)$ and $Q^{(2)}(3,5)$.

For $\theta_-\geq1$, let $R_0$ be the initial up set and define
\begin{equation}
R_{j+1}=R_j\cup\bigl\{i:\ \sigma_i(0)=-1,\ \#\{l\sim i:\ l\in R_j\}\geq\theta_-\bigr\}.
\label{eq:bootstrap}
\end{equation}
The closure is $\bigcup_{j\geq0}R_j$. The closure generation $g(i)$ is the first $j$ for which $i\in R_j$, with $g=0$ on $R_0$ and $g(i)=\infty$ outside the closure.

The closure constrains first up flips even when up spins can later flip back down. We order the initially down sites by the times of their first up flips. When site $i$ first flips up, it has at least $\theta_-$ up neighbors, each of which belongs to $R_0$ or is an initially down site that first flipped up earlier. By induction on first-flip times, we assume that all such earlier sites belong to the closure and let $j_*$ be the largest generation among the up neighbors of $i$. Then $i$ has at least $\theta_-$ neighbors in $R_{j_*}$, so $g(i)\leq j_*+1$. Sites outside the closure therefore never flip up. Tracing back from $i$, each time to the up neighbor with the largest generation, the first-flip times decrease strictly and the generation drops by at most one per step, so this backward chain ends in $R_0$ and contains at least $g(i)$ first flips in sequence, including that of $i$. The closure gives only a necessary condition, and a finite $g(i)$ does not guarantee that site $i$ flips up.

Let $t_1(i)$ denote the first up-flip time of an initially down site $i$, in sweeps, and let $r_i$ be its graph distance to $R_0$. For any $c>0$,
\begin{equation}
\Pr\bigl(t_1(i)<c\,r_i\bigr)\leq(\mathrm e\,zc)^{r_i}.
\label{eq:linear}
\end{equation}
To prove this bound, write $r=r_i$. Since the backward chain moves between nearest neighbors and ends in $R_0$, its last $r$ first flips occur at sites $j_1,\ldots,j_r=i$ in which consecutive entries are nearest neighbors and which are selected in this order in the attempt sequence. For fixed $i$, there are at most $z^{r-1}$ such site sequences. For a fixed sequence, the probability that the first $K$ attempts select its sites in order is at most $\binom{K}{r}N^{-r}\leq\bigl(\mathrm eK/(rN)\bigr)^r$. Taking $K=\lfloor crN\rfloor$ and a union bound over all sequences gives $\Pr(t_1(i)<cr)\leq z^{r-1}(\mathrm ec)^r\leq(\mathrm e\,zc)^r$. Choosing $c=1/(2\mathrm e\,z)$ gives $\langle t_1(i)\rangle\geq(1-2^{-r_i})\,r_i/(2\mathrm e\,z)$. Absorption in $\bs_+$ requires every initially down site to flip up at least once. For the stripe preparation B on the square lattice, the central down sites lie at distance $L/8$ from $R_0$, so $\langle T_B\rangle\geq(1-2^{-L/8})\,L/(64\mathrm e)$ grows at least linearly with $L$.

For the one-dimensional configurations of Eq.~\eqref{eq:exact1d}, $Q^{(1)}(0,2)$ does not allow up spins to flip down. Every down site of A always touches an up neighbor, so when $r$ down sites remain, each attempt removes one of them with probability $r/L$. The single down domain of B has exactly two flippable end sites while $r\geq2$ sites remain, with success probability $2/L$, and the last site has success probability $1/L$. The waiting times of the stages are independent and geometrically distributed, and writing $K_A$ and $K_B$ for the numbers of attempts until absorption, their probability generating functions are
\begin{equation}
\begin{aligned}
\mathbb E[u^{K_A}]&=\prod_{r=1}^{L/2}
\frac{(r/L)u}{1-(1-r/L)u},\\
\mathbb E[u^{K_B}]&=
\left[\frac{(2/L)u}{1-(1-2/L)u}\right]^{L/4-1}
\frac{(1/L)u}{1-(1-1/L)u}.
\end{aligned}
\label{eq:C2}
\end{equation}
Adding the mean waiting times of the stages and dividing by $L$ gives Eq.~\eqref{eq:exact1d}, and the slowest stage, with success probability $1/L$, is common to both and gives the shared tail factor of Sec.~\ref{sec:mpemba}. For each increase of $L$ by 4, the mean time of B increases by $1/2$, while that of A increases by $1/(L/2+1)+1/(L/2+2)<1/2$. Since $H_8>5/2$ at $L=16$ and $H_{10}<3$ at $L=20$, B arrives sooner for $L\leq16$ and A arrives sooner for all $L\geq20$.

Under $Q^{(G)}(0,z+1)$ on a finite $z$-regular graph with $N$ sites, up spins never flip and every down spin flips up the first time it is selected. When $r$ down sites remain, each attempt removes one with probability $r/N$, so the absorption law is the first line of Eq.~\eqref{eq:C2} with $L$ replaced by $N$ and $L/2$ by the initial number $m$ of down sites. It does not depend on how the down sites are arranged, and its mean is $H_m$ sweeps. This gives Eq.~\eqref{eq:exact1d-bulk} for $Q^{(1)}(0,3)$ and the stripe results for $Q^{(2)}(0,5)$.

Equal magnetization and energy do not fix the mean absorption time. For $Q^{(1)}(0,2)$, let the vector $\boldsymbol\ell$ collect the lengths of the down domains, let $a(\ell)=\min(2,\ell)$ count the flippable sites of a domain, and let $F(\boldsymbol\ell)$ be the mean absorption time. The first-step equation is
\begin{equation}
F(\boldsymbol\ell)=
\frac{1+\sum_{j:\ell_j>0}a(\ell_j)F(\boldsymbol\ell-\boldsymbol e_j)}
{\sum_{j:\ell_j>0}a(\ell_j)},\qquad F(\boldsymbol0)=0.
\label{eq:C3}
\end{equation}
On a periodic chain with $L=32$, the down-domain lengths $(8,8)$ and $(14,2)$ have the same numbers of down sites and domains, and hence the same magnetization and energy, yet their mean absorption times are $5.4151$ and $7.5051$ sweeps, respectively.

In the simulations of Figs.~\ref{fig:stripes} and \ref{fig:coarsening}, every trajectory reaches a fixed point and none is discarded. In Fig.~\ref{fig:coarsening}(b), the preparation times are $t_{\rm pre}=0,1,2,4,\ldots,512$, and copies of each prepared state evolve under both $Q^{(2)}(3,4)$ and $Q^{(2)}(3,5)$, so the two operators act on the same preparations.

\end{document}